\documentclass[twocolumn]{aa}
\usepackage{graphicx}
\usepackage{hyperref}

\begin{document}
%
 

\def\Msun{{~\rm M}_\odot}
\def\gsim{\!\!\!\phantom{\ge}\smash{\buildrel{}\over
  {\lower2.5dd\hbox{$\buildrel{\lower2dd\hbox{$\displaystyle>$}}\over
                               \sim$}}}\,\,}
\def\lsim{\!\!\!\phantom{\le}\smash{\buildrel{}\over
  {\lower2.5dd\hbox{$\buildrel{\lower2dd\hbox{$\displaystyle<$}}\over
                               \sim$}}}\,\,}
\def\kms{\rm ~km~s^{-1}}

\def\js{J$_{\rm s} $}
\def\ks{K$_{\rm s} $}
\def\hst{{\sl HST\/}}
\def\chandra{{\sl Chandra\/}}


\title{The Crab pulsar and its red knot in the near-infrared\thanks{Based on 
observations collected at the European Southern Observatory, 
Paranal, Chile (ESO Programme 66.D-0251).}}

\author{Jesper Sollerman}

\institute{Stockholm Observatory, AlbaNova,
SE-106 91 Stockholm, Sweden}

\offprints{jesper@astro.su.se}

\abstract{We present near-infrared observations obtained with ISAAC on the VLT
of the Crab pulsar and its environment. Photometry of the pulsar in \js, 
H and \ks~shows the pulsar spectrum to extend fairly smoothly from the UV/optical
regime. PSF subtraction of the pulsar allows us to study its immediate
neighborhood in some detail. In particular, the knot positioned 
just $0\farcs6$ from the pulsar has been revealed in the IR. Using also archival 
HST data for the knot, we have measured its broad band spectrum to rise steeply 
into the IR, in contrast to the spectrum of the pulsar itself.
}  

\date{Received: Accepted:}

\titlerunning{The Crab with ISAAC}
\authorrunning {Jesper Sollerman}

\maketitle

\section{Introduction}

The Crab nebula is the remnant of a supernova that exploded in 
1054 AD (see e.g., Sollerman et al. 2001 and references therein). 
At the heart of the nebula resides the Crab pulsar, the 
m$_{V}\sim16$ object that powers the entire visible nebula. 
Although the Crab nebula and its pulsar are among the most studied objects 
in the sky, this astrophysical laboratory still holds many secrets about
how supernovae explode and about how pulsars radiate and energize their
surrounding nebulae.

A main theme for pulsar research has been to understand the emission mechanism
for the non-thermal pulsar radiation. This is still to be accomplished. 
No comprehensive model exists 
that can explain all the observed features 
of the radiation. Observationally, only recently was
a broad range UV-optical spectrum of the Crab pulsar published (Sollerman et
al. 2000). Here we have extended this study of the pulsar spectral energy distribution into the near-infrared (IR).

Even if most of the research on the Crab pulsar has concerned the
radiation mechanism, almost all of the spin-down energy comes
out in the particle wind. The stunning image of the pulsar environment 
obtained with \chandra\  captures a glimpse of the energetic processes at work
(Weisskopf et al. 2000).

Direct evidence of the pulsar activity has long been seen in the system
of moving synchrotron wisps close to the pulsar itself. 
The detailed study of the wisps was initiated by Scargle (1969), and is now 
conducted at higher resolution using the \hst\ (Hester et al. 1995; Hester et al. 2002).

Perhaps the most astounding
discovery in these \hst\ images was the knot situated merely $0\farcs6$ 
from the pulsar. At 2~kpc this amounts to a
projected distance of only 1000~AU. Hester~et~al.~(1995)
interpreted this feature as a shock in the pulsar polar wind.
Our IR observations allow us to  further investigate
these manifestations of the magnetic relativistic wind from the pulsar.

\begin{figure*}[t]
\setlength{\unitlength}{1mm}

\begin{picture}(180,120)(0,0)

\put (0, 0)   {\includegraphics[width=59mm,bb=43 43 362.36 364,clip]{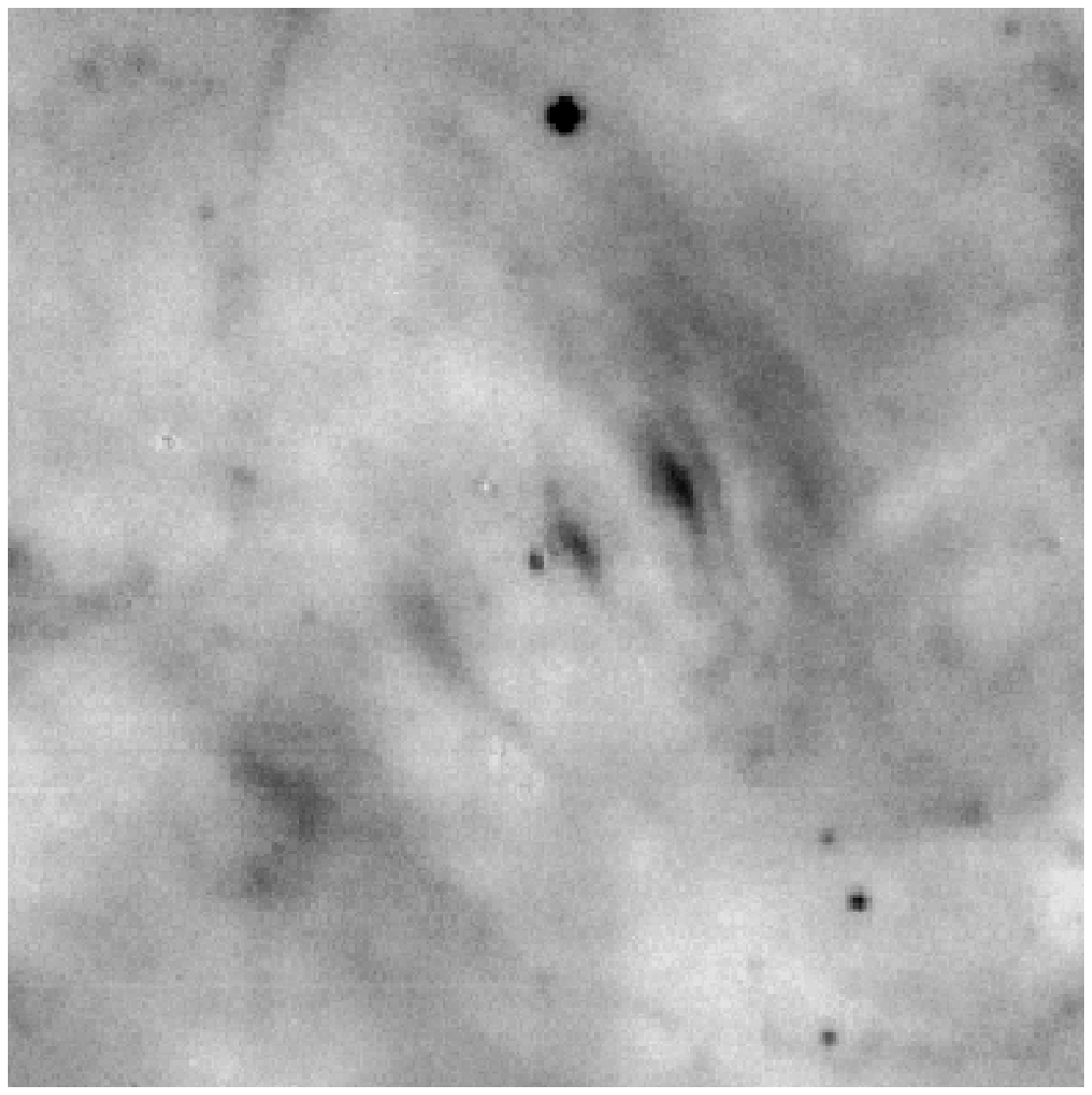}}
\put (60, 0)   {\includegraphics[width=59mm,bb=43 43 362.36 364,clip]{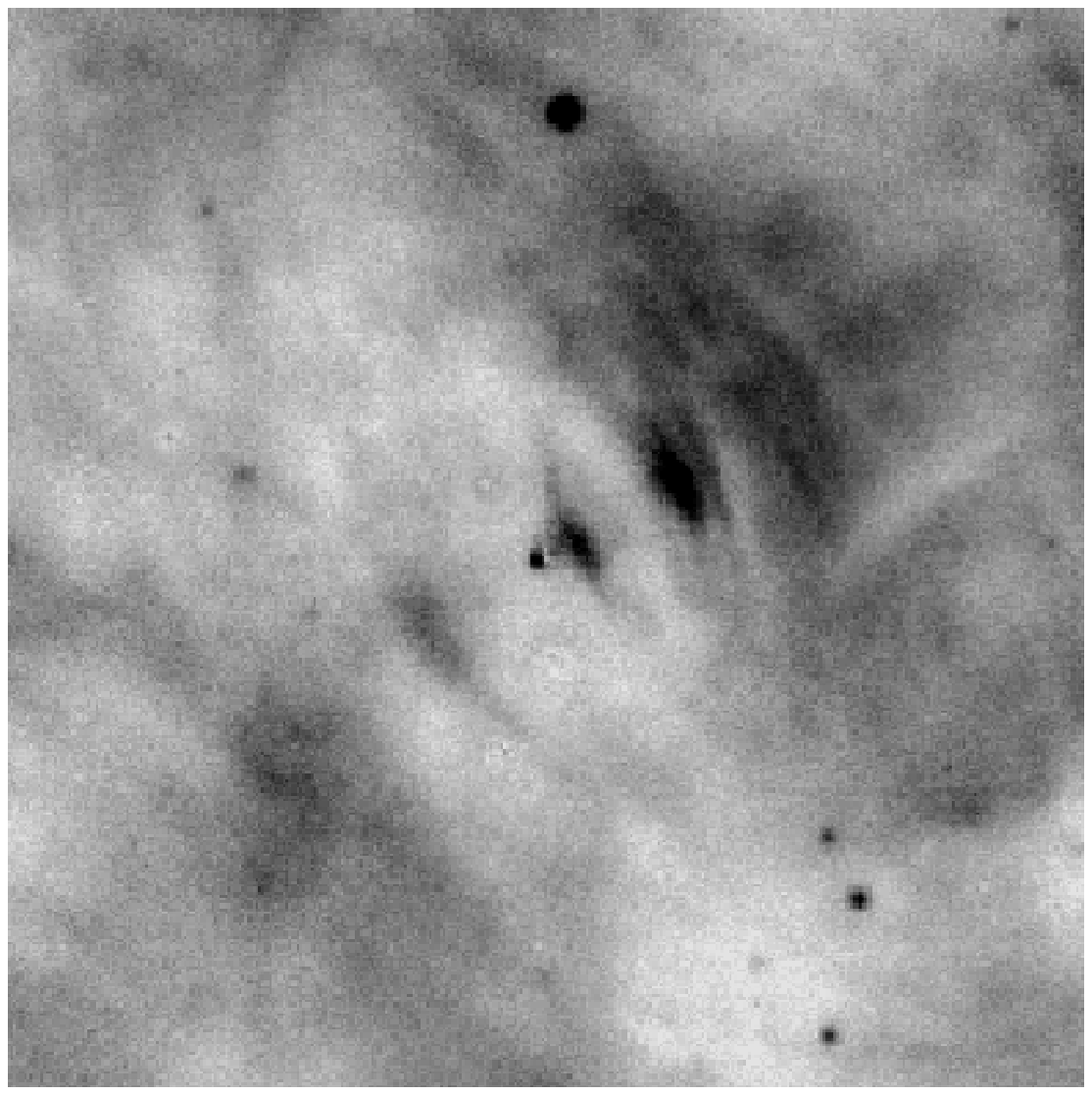}}
\put (120, 0)   {\includegraphics[width=59mm,bb=43 43 362.36 364,clip]{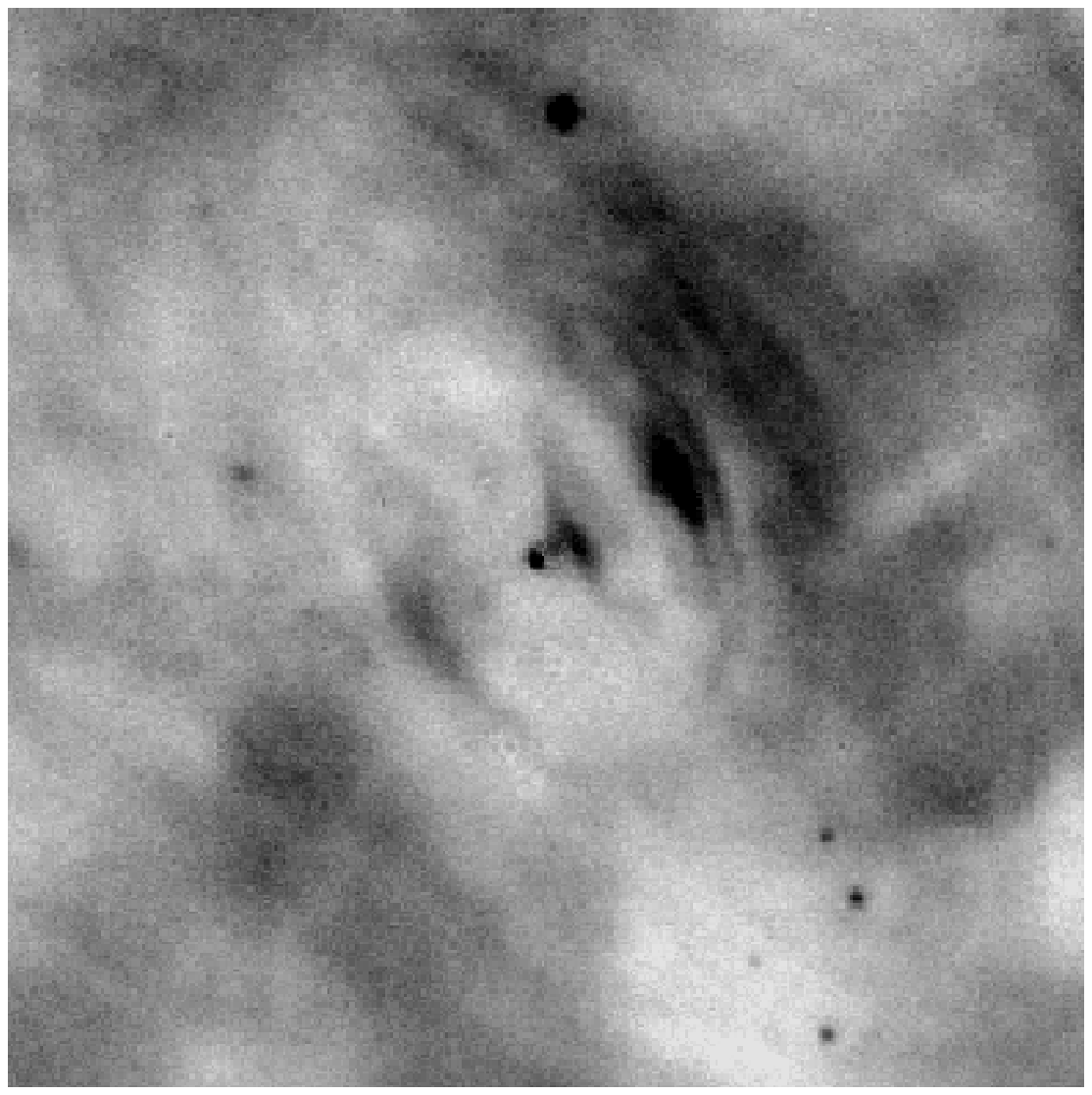}}
\put (0, 60)   {\includegraphics[width=59mm,bb=43 43 362.36 364,clip]{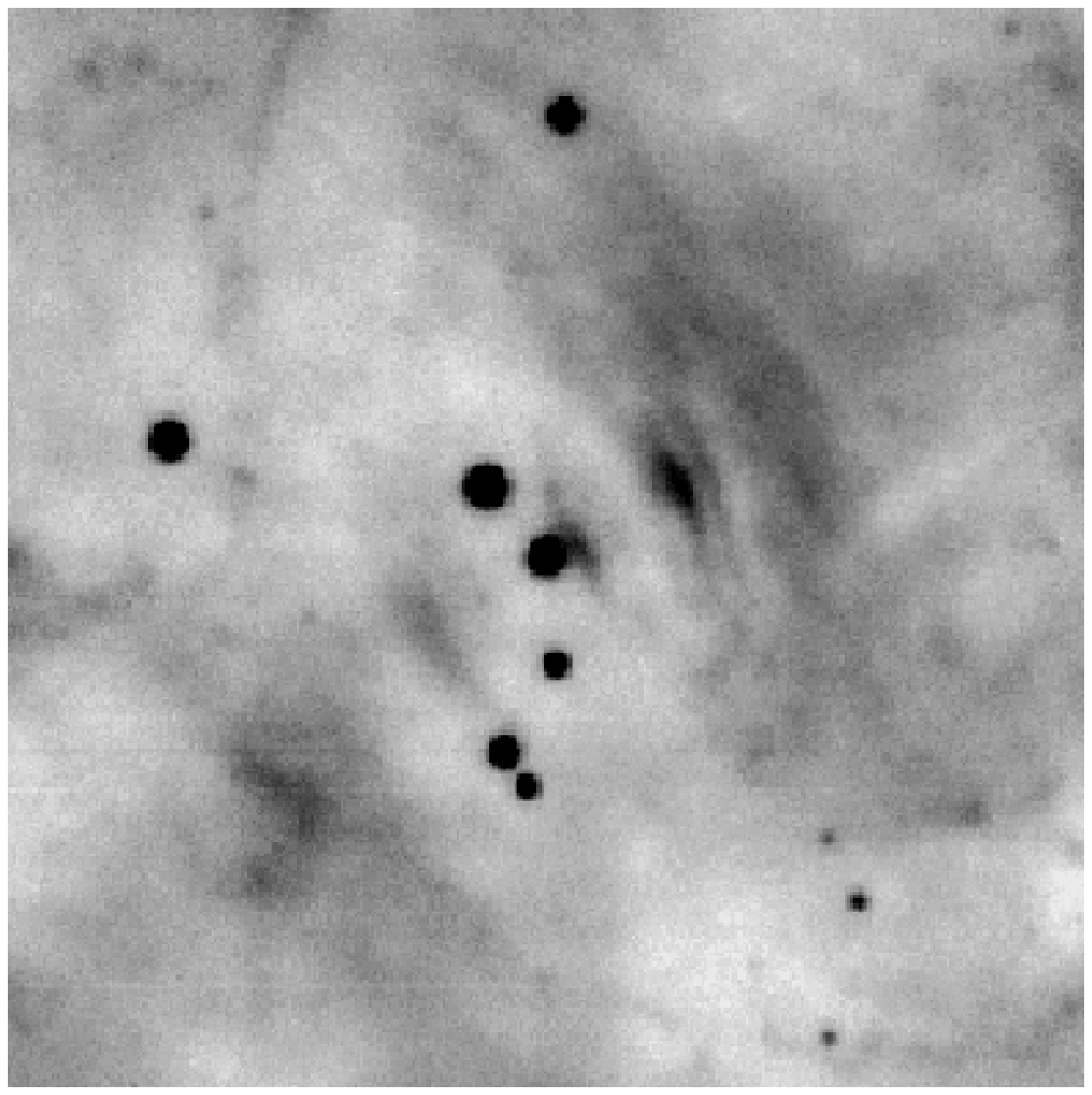}}
\put (60, 60)   {\includegraphics[width=59mm,bb=43 43 362.36 364,clip]{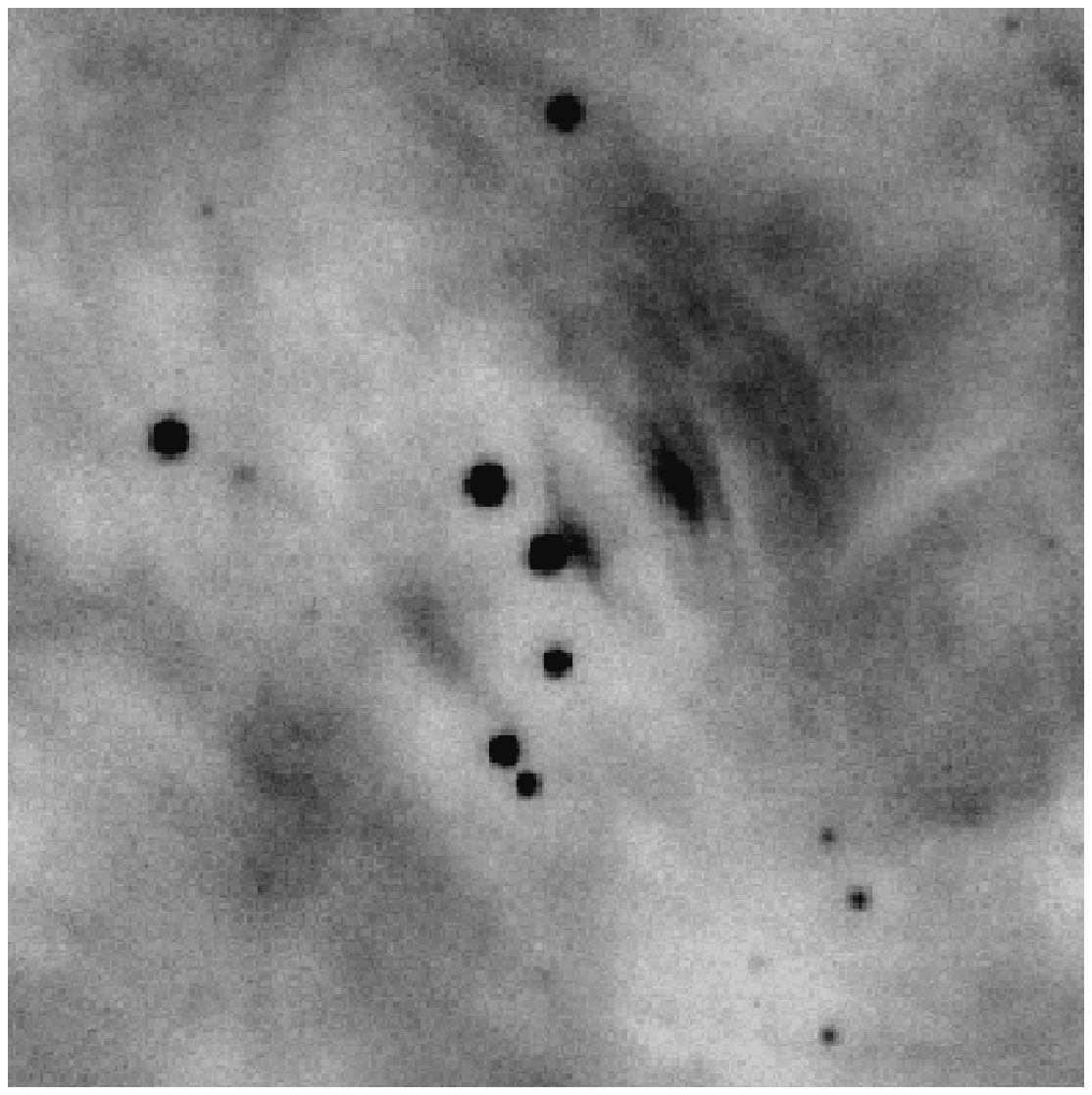}}
\put (120, 60)   {\includegraphics[width=59mm,bb=43 43 362.36 364,clip]{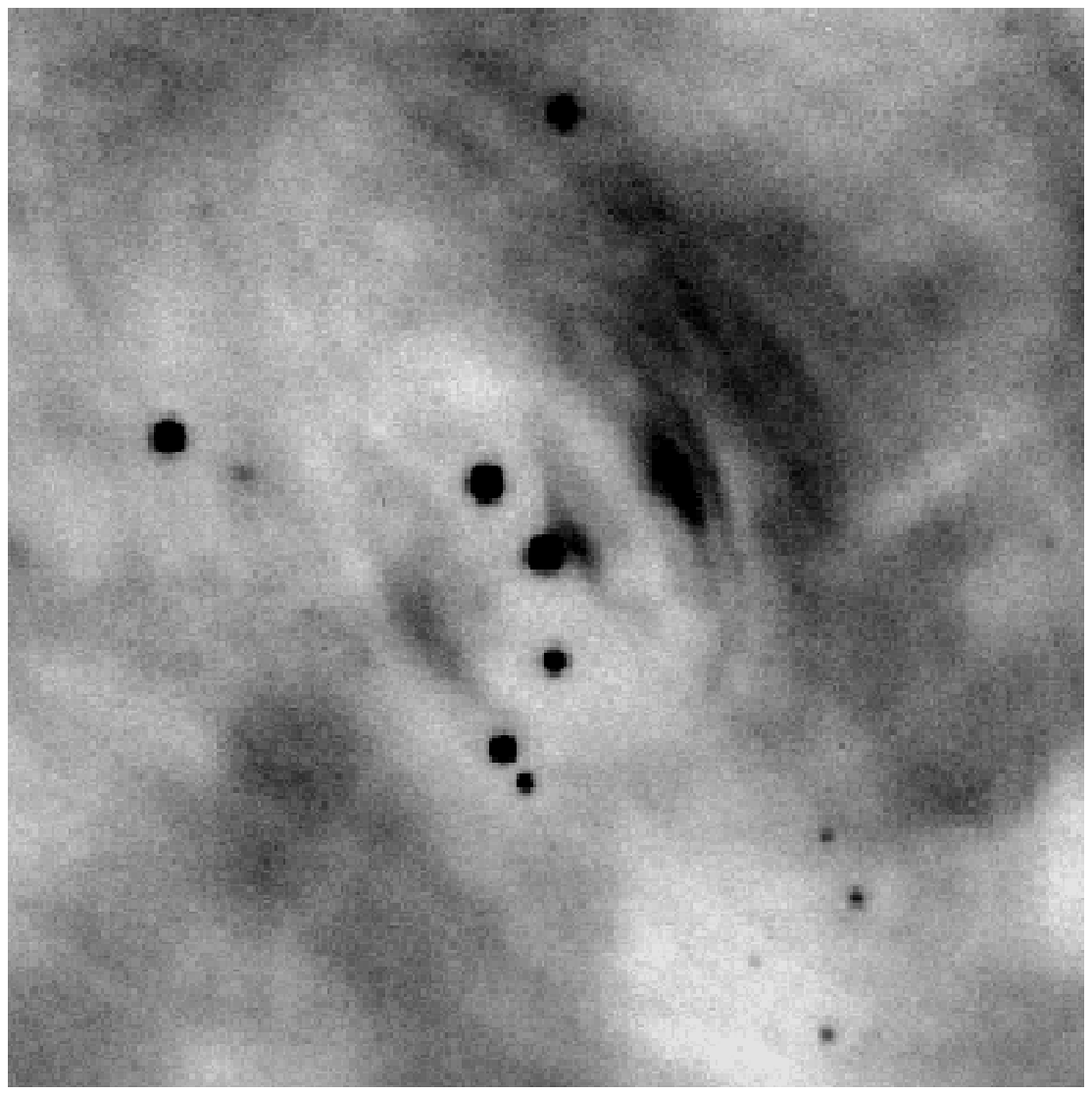}}

\end{picture}
\caption{The central part of the Crab nebula in the infrared, \js\ (left), H (middle) and \ks\ 
(right).
Observations obtained with ISAAC on 13 October 2000. 
The pulsar is the lower right 
(South Preceding) of the two bright objects near the center of the field.
The field of view shown is $60\arcsec\times60\arcsec$ centered on the pulsar. North is up and East to the left. The upper panels show the original images, and the lower panels show the images where several objects, including the pulsar, have been PSF subtracted. The lower panels are intended to reveal the knot positioned very close to the pulsar. Detailed color images of these frames 
are presented in Sollerman \& Flyckt (2002).
}
\end{figure*}

\section{IR photometry, reductions and results}

\subsection{Observations and basic reductions}

We obtained IR imaging in the short wavelength (SW) mode of 
ISAAC\footnote{Infrared Spectrometer And Array Camera, See
\href{http://www.eso.org/instruments/isaac/ for details}
{http://www.eso.org/instruments/isaac/}} at UT1 
of VLT on October 13, 2000. 
ISAAC (Moorwood et al. 1999) is equipped with a $1024\times1024$ pixel Rockwell
Hawaii array providing a plate scale of $0\farcs147$~pixel$^{-1}$
and a total field of view of about $2\farcm5\times2\farcm5$.
The exposures were obtained in \js, H and \ks. 
The \js\ filter is a narrower version of the classical J band filter. It is centered on $1.24\mu$m 
with a width of $0.16\mu$m, and is 
constructed to
deliver more accurate photometry due to a significantly lower background without any real loss in
efficiency (e.g., Simons \& Tokunaga 2002). 
As the pulsar is a relatively bright source, short exposure times were
used. In all cases the observations were obtained using a 
detector integration time of 2.0 seconds. For each position 13 such exposures 
were stacked before readout, and we obtained images at 6 positions with the 
telescope being offset by random amounts up to 25 arcsec between the individual
images.
The total exposure time was thus 156 seconds per band. 
The excellent image quality provided by ISAAC also allowed a detailed
view of the central region of the Crab nebula.
The images are displayed in Fig. 1.

The data were reduced using both $\tt IRAF$ and the $\tt JITTER$ routine within
the $\tt ECLIPSE$ package. This includes bias subtraction, flat fielding,
sky subtraction and combination of the different exposures.  We found
no significant difference in the photometry obtained from these reductions packages.

The photometry was achieved in the following way.  
Photometric standard stars are observed regularly with ISAAC in service mode.  
On  October 13 the stars S889-E and S121-E (Persson et al. 1998)  were
observed in J, \js, H and \ks. S121-E was observed immediately
before the Crab observations. We reduced these images and measured
the flux of the standard stars to establish zero-points in all filter bands.  
This night had very low humidity and the seeing was typically 
just under $0\farcs8$. The FWHM varies from
$\sim 0\farcs88-0\farcs65$.
in the final \js\ and \ks\ images,
respectively. 
No atmospheric extinction corrections were applied, as they are
smaller than 0.01 mag.

\subsection{Photometry}

Photometry was obtained of the Crab pulsar and some of the stars in
the field using point-spread function (PSF) fitting. 
A PSF was constructed
using bright and relatively isolated stars in the frame.
This PSF was then
subtracted from the image of the pulsar and some nearby stars 
using $\tt ALLSTAR$
within the $\tt DAOPHOT$ package
(Stetson 1987). 
This improves the accuracy in this crowded region, 
in particular for the pulsar
which has a complex surrounding.
The magnitudes were aperture corrected using bright isolated standard stars.
The aperture corrections amounted to about 0.1 mag. 
For the three isolated aperture stars we measure no significant offset 
between our photometry and the photometry in the 2MASS catalogue. 
They agree to within 0.05 mag.
For the more crowded and fainter stars we regard our photometry to be
superior, due to the excellent seeing and to the performed PSF subtraction.
This is particularly true for the Crab pulsar.
The magnitudes for the Crab pulsar, and several of its nearby stars,
are presented in Table~1.

\begin{table}[b]
\caption[]{\js, H and \ks~photometry of the Crab pulsar and its 
nearest neighbors$^{a}$}
\label{table}
\begin{tabular}{lccc}
\hline
Offsets from Pulsar & \js\ & H & \ks\ \\
 (arcsec) & & & \\

\noalign{\smallskip}
\hline
\noalign{\smallskip}

Crab Pulsar~& 14.80 &  14.27 & 13.77 \\
6.29~~N,~~20.37~E & 14.64 &  14.26 & 14.14 \\
3.82~~N,~~3.24~~E & 14.08 &  13.75 & 13.65 \\
5.84~~S,~~0.51~~W & 16.14 &  15.41 & 15.24 \\
10.58~S,~~2.29~~E & 15.29 &  14.90 & 14.80 \\
12.43~S,~~1.05~~E & 16.50 &  16.11 & 16.06 \\

\hline
\end{tabular} \\
\begin{tabular}{lll}
$^a$ \ The errors $\lsim0.05$ mag in all bands && \\
\end{tabular}
\end{table}

The formal errors for the pulsar and for the nearby stars 
from $\tt DAOPHOT$ are less than 0.01 mag.
The uncertainties in the standard star magnitudes are 
also much below 0.01 mag, and
the zeropoints obtained for the two standards agreed to within 
0.01 mag in all bands.  
The total error is instead likely to be dominated by small errors in the 
PSF subtraction and in the aperture corrections. 
We estimate a conservative error budget of 0.05 mag for the 
pulsar and the nearby stars in the three bands observed.

Using the conversions from Wilson et al. (1972) we have converted the
magnitudes to fluxes. Those were then dereddened using $E(B-V)$=0.52
(Sollerman et al. 2000), and the extinction law from Fitzpatrick
(1999). The results are plotted in Fig. 2, which also includes the
optical-UV data from Sollerman et al. (2000).

\begin{figure}
\begin{center}
\includegraphics[width=.5\textwidth]{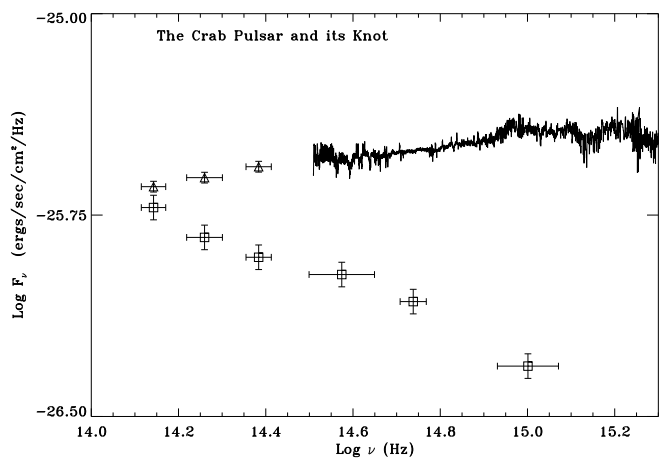}
\end{center}
\caption[]{Spectral energy distribution of the Crab pulsar. 
The optical and UV data are from Sollerman et al. (2000). 
The triangles are our new 
ISAAC measurements. Also shown (squares) are the fluxes of the knot, 
here multiplied by a factor ten. The optical data for the knot is from \hst. 
The knot data error bars illustrate an uncertainty of $10\%$ in the flux.
All observed fluxes were dereddened using R=3.1 and $E(B-V)$=0.52. 
}
\label{spec2}
\end{figure}

We note that our IR magnitudes deviate from the recent
results in the 2MASS point source catalogue. Our measurements give a pulsar 
that is fainter by 0.32, 0.35, and 0.27 mag in \js, H and \ks, 
respectively. The 2MASS magnitudes are consistent with the data from
Eikenberry et al. (1997) (to within 0.1 mag), but since 
our relatively isolated standards in the field show
good agreement with 2MASS we do not think this is due to an offset in
the zero-point. 

Our ISAAC photometry has excellent signal and image quality. 
In the crowded region around
the pulsar, this significantly improves the background subtraction. 
In particular, PSF subtraction excludes contributions from wisps and the
nearby knot (Fig.~1). The accuracy of the PSF fit can be seen in the 
subtracted images, where the knot is clearly revealed.
We believe that the difference in magnitudes is due to our ability to better separate the
background from the pulsar emission.

We note that our measurements show a good agreement with previous
measurements using time-resolved photometry of the Crab (Penny 1982;
Ransom et al. 1994), although the errors claimed by Ransom et al. are rather large.
Such measurements generally use a large aperture
and simply assume any non-varying contribution to be due to the
background. By integrating under the pulsar light curve, they measure
the pulsating contribution of the pulsar flux.  
We believe that with the good image quality of our data, we are now able to
subtract virtually all background from the pulsar contribution.
The implications of the pulsar photometry are discussed below.

\subsection{The IR images}

Our new images show most of the well known features of the inner part of the 
Crab Nebula. 
In particular, the wisps are clearly visible in higher detail than
previously obtained in the IR. Some filaments are also seen, most
strongly in the \js\ band. This is most likely due to emission from 
[Fe~II]~$1.26\mu$m.  The [Fe~II]~$1.64\mu$m line in the H-band 
was detected by Graham et al.~(1990). It appears less conspicuous, probably 
also because
the narrower \js\ filter includes less continuum. 
Finally, the \ks\
band image is clearly dominated by amorphous synchrotron emission.
With suitable cuts the pulsar image appears
slightly elongated. This is due to the presence of the knot first
identified by Hester et al. (1995) on \hst/WFPC2 images.  
To reveal this structure in our images we constructed a 
PSF of suitable stars in the image, and subtracted this from the image of
the pulsar and some nearby stars.
In the subtracted images, the knot is clearly visible in all 
three bands (Fig.~1).

From our \js, H and \ks\ band 
images it is possible to create a color image. 
This was done by Sollerman \& Flyckt (2002), who also
presented a color image made out of the PSF 
subtracted frames. In the latter image, 
it can be seen that the knot is redder than, e.g., the wisps.

To quantify this color difference 
we have estimated the magnitudes of the knot as
well as of the nearby wisp 1 (Scargle 1969).  The knot
was measured within an aperture of $0\farcs9$ and the wisp was also simply
measured with a circular aperture with a radius of $1\farcs2$.  
The spectral energy distribution of the knot is shown 
in Fig.~2. It is clearly red. In the \ks-band the flux from the knot 
amounts to about $8\%$ of the flux of the pulsar.
The stationary wisp appears to have a flatter spectrum in this regime, 
as does the pulsar itself (see below).

\section{Optical data from the Hubble Space Telescope}

Data on the Crab pulsar are also available in the \hst\
archive.  Most of the observations are from the
comprehensive monitoring
programme of the inner parts of the nebula 
(P.I. J. Hester).  These frames allow a detailed study of both the
spectral and temporal  properties of the knot. In August 1995 the
region was observed in  three filters (F300W, F574M, F814W).  We
performed aperture photometry of the knot at these images. An aperture
with a radius of 7 pixels (0\farcs32) was used, and the background was
measured in an equal size aperture positioned at the opposite side of
the pulsar. The magnitudes were converted to dereddened
fluxes and are plotted in Fig.~2. The uncertainties in these fluxes
are not negligible.  This is because the pulsar is saturated in all
images, and no accurate PSF subtraction of the pulsar could be
made. 
The absolute flux of the knot also depends on the chosen aperture size. 
However, to estimate a spectral energy distribution we measured the knot with the same aperture size 
in the three filter images. Then the formal errors due to photon statistics is $\sim5\%$, but 
the position of the region where we measure the background can introduce uncertainties of up to 
$\sim10\%$. In Fig.~2 we have included error bars of $10\%$ in the flux for all knot measurements, 
and also indicated the widths of the used filters.
Despite these uncertainties, the color trend found is unambiguous and confirms
the results from our IR measurements.  We measured the dereddened
spectral index for the knot to be $\alpha_{\nu}\sim-0.8$ in the \hst\ 
bands  ($F_{\nu} \propto \nu^{\alpha_{\nu}}$)  in full agreement
with the slope determined from the IR data (Fig. 2).  The spectral
energy distribution of the wisp was also estimated in both the \hst\ and
ISAAC images, and the slope was found to be negative ($\alpha_{\nu}<0$) 
although substantially flatter than for the knot. The knot is indeed redder
than the surrounding nebula, and substantially redder than the pulsar
itself.

The wealth of data in the F574M filter available in the archive  also
allows a study of the temporal behavior. The knot is present  in all
frames, and thus appears quasistationary for more than six years,
although the position appears to vary at  the $0\farcs1$ level. The
dereddened flux of the knot within the aperture  is measured to be
$\sim9\times10^{-28}$~ergs~s$^{-1}$~cm$^{-2}$~Hz$^{-1}$ in August 1995,  
but variations of the flux by at least $50\%$ are observed. This means
that the rather smooth connection between the optical and IR data for
the knot in Fig.~2 may be partially superficial.  However, we have
also checked the F547M data from 1995 against \hst\ data taken the
same day as our ISAAC data. We measured only a small increase
($\sim10\%$) in the flux of the knot between the 1995 image and the image
obtained in 2000. This is likely to be
within the errors of our measurements and no correction for this
evolution has been applied to Fig.~2.

\section{Discussion and Implications}

\subsection{The Pulsar}

We have reported magnitudes for the Crab pulsar in the IR,
where the observational situation has  previously been rather
uncertain (compare, e.g., estimates by Penny 1982; 
Middleditch et al. 1983;  Eikenberry et al. 1997). Together with recent
optical-UV data this significantly revises the observational basis for
the pulsar emission mechanism.  The slope in the near-IR is somewhat
steeper than in the UV/optical region.  A power-law fit to the three
near-IR points gives a slope of $\alpha_{\nu}=0.31\pm0.02$, whereas
the  UV/optical spectral index is  $\alpha_{\nu}=0.11$  (Sollerman et
al. 2000).

A spectral index of one third is expected if the nonthermal emission
is due to synchrotron radiation   of relativistic particles in the
magnetosphere  of the pulsar.  For a monochromatic particle
distribution  a spectral flux of $F_{\nu} \propto \nu^{1/3}$ is
expected in the low frequency range below some maximum frequency
$\nu_m$ (e.g., Ginzburg \& Syrovatskii 1965).

Models for the optical emission from the Crab pulsar
based on the
synchrotron mechanism often aim at reproducing the maximum frequency
in the optical range 
($\nu_m\sim5\times10^{14}$ Hz, eg.,  Stoneham 1981; Mavlov \&
Machabeli 2001; Lyne \& Graham-Smith 1998).
This is mainly based on the observations by
Oke (1969), and is 
not supported by the more recent observations presented in Sollerman et
al. (2000). A model in which the flat spectrum extending all the way
into the UV is due to synchrotron emission at very small pitch angles
(Crusius-W\"atzel et al. 2001) assumes a very
rapid spectral decline towards infrared frequencies. This has been reported by
Middleditch et al. (1983) and is sometimes interpreted as synchrotron
self-absorption (see Lyne \& Graham-Smith 1998 and references
therein). The near-IR data presented here do not support such a steep 
decline, rather a possible smooth leveling of the spectral index from UV via 
the optical and into the near-IR.
If the L-band measurement of Penny (1982) is
correct, as seems to be the case for his J, H, and K band data, there is no
evidence for a sharp decrease into the L-band.  New observations are
underway to investigate the IR spectral shape (see below).
These new data call for a fresh look on the emission mechanism scenarios for 
young pulsars.

\subsection{The Knot}

For the knot, we have shown that its structure is quasi-stationary, 
and that its emission has a red spectrum. 
Most quantitative modeling for the overall properties of the Crab nebula 
is based on the theory outlined by Rees \& Gunn (1974) and extended by
Kennel \& Coroniti (1984). However, few 
papers address the properties of the knot.
Lou (1998) presented a formation scenario in terms of MHD theory, 
while Shapakidze \& Machabeli (1999) argue for a plasma mechanism. None 
of these scenarios predict a very red spectral distribution.

Another area where caution may be required is in the   recent claims
of weak off-pulse emission from the Crab pulsar in the visible
(Golden et al. 2000).  It is clear that a knot close to
the pulsar has to be seriously  considered in these kinds of
studies. The off-pulse emission detected by Golden et al. (2000) could
either be from the pulsar itself, or from a nearby knot. The strong
polarization and in particular the red spectrum of the unpulsed
emission, in contrast to the pulsed pulsar emission, is clearly consistent
with a knot. The knot we have detected in the IR is 0\farcs6 from the
pulsar and could probably be excluded if the seeing is good enough and
the neighboring star is used as a PSF star, but more nearby features
may also be present. It is also unclear if the knot 0\farcs6 from the
pulsar is luminous enough to explain the off-pulse emission discussed
in Golden et al. (2000; Shearer \& Golden 2002), although the
intensity of the knot is clearly varying (see also Hester et
al. 2002).  This at least calls for caution in the interpretation of
the off-pulse data.

\subsection{Other pulsars}

Although the Crab pulsar is the brightest of the optical pulsars, a
few more pulsars have now been detected in the near-IR. A
comparison is made in Fig.~3, adapted from Shibanov et al. (2003).  
It is clear that none of these pulsars show a low-energy break.  In
particular, the previous indication of a break in the Vela pulsar
spectrum at 6500~\AA~(Nasuti et al. 1997) is not supported by
these new observations. The Vela pulsar spectrum is rather flat over
the entire optical/IR regime. Using the best available data;
ground-based UB photometry from Nasuti et al. (1997), \hst\ F555W data
from Mignani \& Caraveo (2001), and \hst\ F675W and F814W  as well as
VLT/ISAAC \js\ and H data from Shibanov et al. (2003), we can fit a
power law to the entire optical/IR regime.  The amount of extinction
is somewhat uncertain, affecting the derived spectral index. Including
the errorbars in the fit gives $\alpha_{\nu}\sim-0.1$ for
A$_{V}$=0.18 and $\alpha_{\nu}\sim+0.1$ for A$_{V}$=0.4.

\begin{figure}
\begin{center}
\includegraphics[width=.4\textwidth]{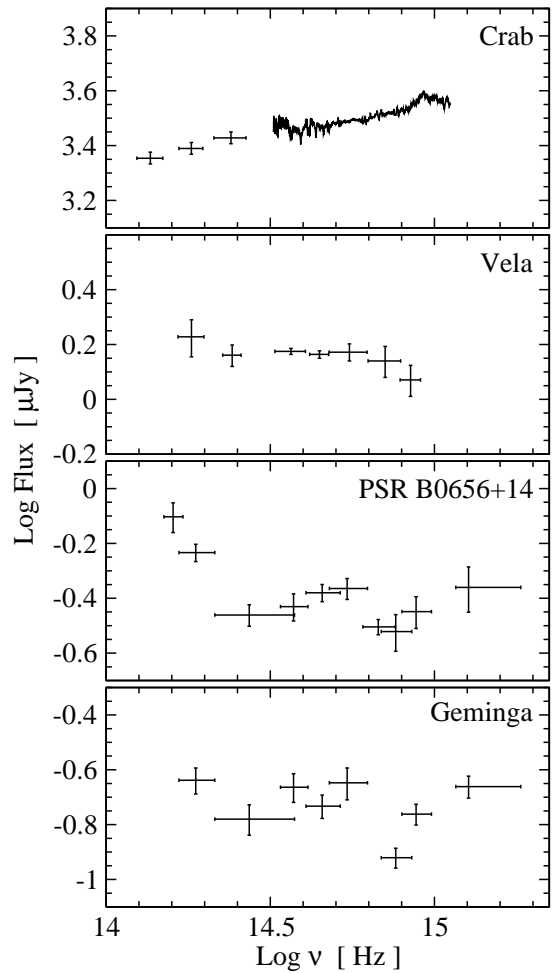}
\end{center}
\caption[]{Comparison between the pulsars with observed near-IR emission, 
from Shibanov et al. (2003). See text for discussion.}
\label{shib}
\end{figure}

As seen in Fig.~3, the middle-aged pulsars Geminga and PSR 0656+14
also show a fairly flat spectrum in this range. The optical emission
can, however, also be interpreted as a combination of two components;
a thermal component and a steeper non-thermal power-law component
which is increasing into the IR (e.g., Pavlov et al. 1997;
Martin et al. 1998). This has been interpreted in terms of 
spectral evolution with pulsar age (e.g., Koptsevich et al. 2001).

In this paper we have demonstrated that accurate measurements of the
magnitudes of the Crab pulsar requires good signal-to-noise and
spatial resolution. It is clear that the uncertainties in measuring
much fainter pulsars can be considerable. It is important to keep
these uncertainties in mind when interpreting the scarce datasets
available for optical pulsars. Many claims in the literature regarding
flux excesses or dips in the spectral energy distribution of optical pulsars
appears to have little significance.

Another interesting result by Shibanov et al. (2003) is the tentative
near-IR  detection of an extended structure around
the Vela pulsar.  This could be an IR counterpart to the X-ray nebula
established by \chandra\ observations (Helfand et al. 2001). As
these structures are not detected in the optical, the spectra must be
very red, in some way resembling the red structures found in the Crab
pulsar neighborhood.

\section{Future plans}

The Crab pulsar and its environment continue to be the prime
astrophysical  laboratory for the study of the pulsar emission
mechanism and the spin-down  powering of pulsar nebulae.  Although
much observational effort has been put into this object,  a modern
re-investigation is likely to clean up the many contradictory
measurements. A project to acquire optical imaging in good seeing is
underway to determine the knot-subtracted spectral energy
distribution of the pulsar. We will also pursue IR observations using the
NAOS/CONICA instrument on VLT to monitor the structures close to the
pulsar in even better spatial resolution. These observations will also
extend into the L-band to clarify if the knot contributes
significantly to the emission at these frequencies, and establish
whether or not the IR drop of the pulsar is real.

\begin{acknowledgements}
Special thanks to Veronica Flyckt who did a substantial part of
the work on this project during her master thesis conducted at ESO Garching.
Also thanks to Yuri Shibanov and Peter Lundqvist for comments on the manuscript, and to 
Bruno Leibundgut
for comments and for support from ESO Office of Science during
Veronicas stay in Garching.
\end{acknowledgements}

\end{document}